\documentclass[%
 aip,
 amsmath,amssymb,
 reprint,%
]{revtex4-2}

\usepackage{graphicx}% Include figure files
\usepackage{dcolumn}% Align table columns on decimal point
\usepackage{bm}% bold math
\usepackage[utf8]{inputenc}
\usepackage[T1]{fontenc}
\usepackage{mathptmx}
\usepackage{xcolor}
\usepackage{soul}
\usepackage{color}
\usepackage{multirow}

\begin{document}

\preprint{AIP/123-QED}

\title{First-principles study of the electronic and optical properties of Ho$_{\rm W}$ impurities in tungsten disulfide}
% Force line breaks with \\
\author{Michael N. Leuenberger}%
 \email{michael.leuenberger@ucf.edu}
\affiliation{ 
NanoScience Technology Center, Department of Physics, and College of Optics and Photonics, University
of Central Florida, Orlando, FL 32826, USA%\\This line break forced with \textbackslash\textbackslash
}%

\author{M. A. Khan}
\email{mahtabahmad.khan@fuuast.edu.pk}
 \affiliation{Department of Applied Physics, Federal Urdu University of Arts, Science and Technology, Islamabad, Pakistan}%Lines break automatically or can be forced with \\

\date{\today}% It is always \today, today,

\begin{abstract}
The electronic and optical properties of single-layer (SL) tungsten disulfide (WS$_2$) in the presence of substitutional Holmium impurities (Ho$_{\rm W}$) are studied. Although Ho is much larger than W, density functional theory (DFT) including spin-orbit coupling is used to show that Ho:SL WS$_2$ is stable. The magnetic moment of the Ho impurity is found to be 4.75$\mu_B$ using spin-dependent DFT. The optical selection rules identified in the optical spectrum match exactly the optical selection rules derived by means of group theory. The presence of neutral Ho$_W$ impurities gives rise to localized impurity states (LIS) with f-orbital character in the band structure. Using the Kubo-Greenwood formula and Kohn-Sham orbitals we obtain atom-like sharp transitions in the in-plane and out-of-plane components of the susceptibility tensor, Im$\chi_{\parallel}$ and Im$\chi_{\perp}$.
The optical resonances are in good agreement with experimental data.
\end{abstract}

\maketitle

\section{Introduction}
Single-layer (SL) transition metal dichalcogenides (TMDs) are very attractive materials because of their special electronic and optical properties that enable lots of promising applications.\cite{review_TMDCS,review_TMDCS_2} Since SL TMDs are semiconductors with a direct band gap,\cite{Direct_Band_Gap_1,Direct_Band_gap_2} 
they can be used to build transistors and optoelectronic devices. Since the band gap is in the visible regime, photodetectors and solar cells can be developed. Growth processes typically introduce defects and impurities in SL TMDs with profound effects on their electronic, optical, and magnetic properties.\cite{Sulfur_vacancies, Zhang_jap, Structural_Defects_Graphen}

Over the past few years we have developed theoretical models based on density functional theory (DFT), tight-binding model, and 2D Dirac equation for the description of the electronic and optical properties of vacancy defects in TMDs,\cite{Erementchouk2015,Khan2017,ER_W_khan} which are naturally occurring during different growth processes, such as mechanical exfoliation (ME), chemical vapor deposition (CVD), and physical vapor deposition (PVD).
 A central result of our papers is that group theory can be used to derive strict selection rules for the optical transitions, which are in excellent agreement with the susceptibility calculated by means of the Kubo-Greenwood formula using the Kohn-Sham orbitals. 
 
 In our recent paper in Ref.~\onlinecite{ER_W_khan} we performed DFT calculations and obtained the optical spectrum of SL WS$_2$ in the presence of substitutional Er$_{\rm W}$ atoms. Although we did not include the effect of spin-orbit coupling (SOC), we obtained good agreement with Bai et al.'s experiments on Er-doped MoS$_2$ thin films using CVD growth\cite{Bai2016} and wafer-scale layered Yb/Er co-doped WSe$_2$.\cite{Bai2018} 
 Similar results have been found by L\'opez-Morales et al.\cite{LOPEZMORALES2021111041}
 One of our motivations was to find out whether some of the LIS of Er lie within the band gap of SL WS$_2$. We were able to show that this is indeed the case. The reason for our motivation is that LIS inside the band gap of a semiconductor can be potentially used as a qubit or a qudit for quantum information processing.
 Remarkably, TMDs with rare-earth atoms (REAs) exhibit the unique property of strong isolation of their electrons in the unfilled 4f shell by the surrounding d shell. This property leads generally to high quantum yields, atom-like narrow bandwidths for optical transitions, long lifetimes, long decoherence times, high photostability, and large Stokes shifts. 
 This strong isolation of the 4f electrons makes them behave like electrons in a free atom. Therefore, it is not surprising that Ce$^{3+}$ impurities in yttrium aluminium garnet (YAG) can reach long coherence times of $T_2=2$ ms.\cite{Siyushev2014}
By replacing YAG with the calcium tungstate CaWO$_4$ as host material, it is possible to avoid the paramagnetic impurities of Y and substantially reduce the nuclear spin concentration without isotopic purification. Consequently, the Hahn echo experiment is able to achieve a long spin coherence time of $T_2=23$ ms for Er$^{3+}$ impurities in CaWO$_4$.\cite{Dantec2021}
 Thus, it is advantageous to identify host materials for REAs with low concentrations or even free of paramagnetic impurities and nuclear spins. We argue here that TMDs are good candidates for such host materials.
 
 \begin{figure*}%[hbt]
	\begin{center}
		\includegraphics[width=7in]{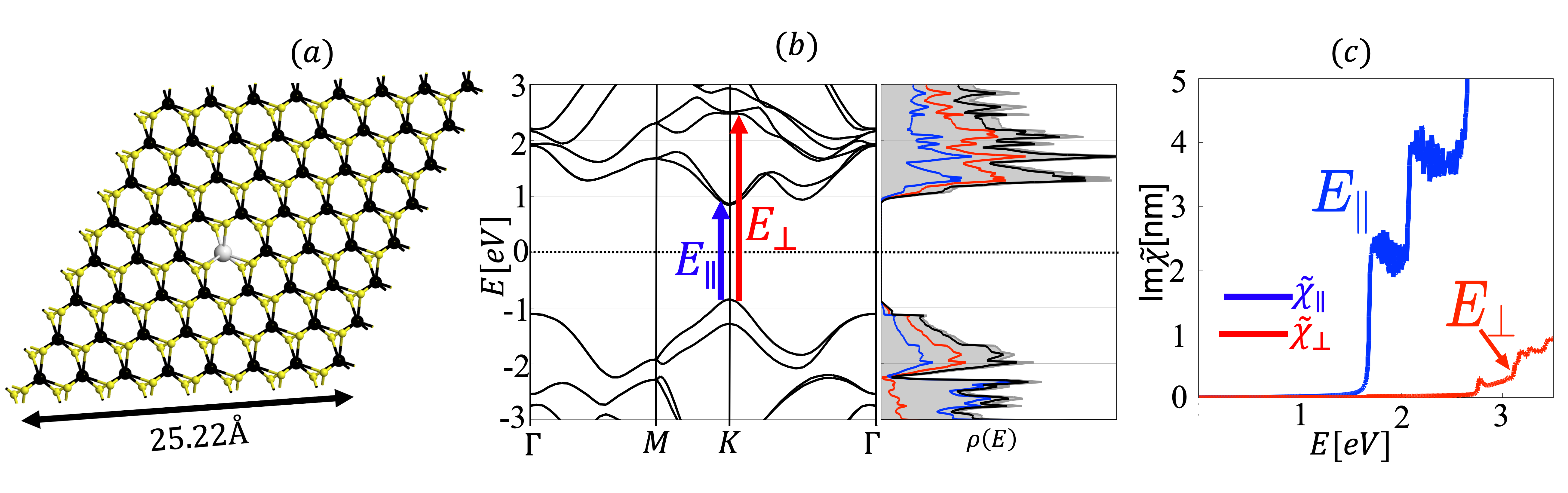}
	\end{center}
	\caption{(a) Schematic shows a Ho$_{\rm W}$ impurity inside a $8\times8\times1$ supercell of SL WS$_2$. The grey circle depicts a Ho atom. The black (yellow) circles represent W (S) atoms. (b) The band structure and density of states (DOS) $\rho(E)$ of pristine SL WS$_2$ exhibits an in-plane band gap of $E_{\parallel}=1.6$ eV  and an out-of-plane band gap of $E_{\perp}=3.2$ eV. 
	%Corresponding irreducible representations (IRs) of valence band (VB) $E^{\prime}$, conduction band (CB) $A^{\prime}_1$ and CB$+1$ $E^{\prime\prime}$ are also shown. 
	States at the valance band edge are splitted due to SOC  with $\Delta_{SOC}=433$ meV. Gray region in the DOS plot specify the total DOS, while red blue and black curves are for $d$-orbitals of W, $p$-orbitals of S and the sum of the contributions, respectively.  (c) Optical response of the pristine WS$_{2}$, showing the in-plane and out-of-plane band gaps.}
	\label{fig:Er_W_defect}
\end{figure*} 

Here, following Ref.~\onlinecite{ER_W_khan}, we calculate the electronic and optical properties of Ho$\rm _W$ impurities in SL WS$_2$. In particular, we find a peak at 2120 nm in the optical spectrum which is in good agreement with the characteristic wavelength of Ho observed in Ho:YAG lasers.\cite{Ho_YAG,Ho_YAG_2,Malinowski2000} Laser systems that operate in the 2 $\mu$m range offer exceptional advantages for free space applications compared to conventional systems that operate at shorter wavelengths. This gives them a great market potential for the use in LIDAR and gas sensing systems and for direct optical communication applications. Besides, we find additional peaks in the optical spectrum of Ho$\rm _W$:SL WS$_2$, which are a direct consequence of the D$_{3h}$ symmetry of the Ho$\rm _W$ impurity and the interplay between the valley angular momentum (VAM), exciton angular momentum (EAM), and lattice angular momentum (LAM).\cite{Momentum_conservation_photon} 

% Pristine TMDs are invariant with respect to the reflection $\sigma_{h}$ about the Mo or W plane of atoms ($z=0$ plane). Therefore, electron states can be classified into two catagories: even and odd or symmetric and antisymmetric with respect to the $z=0$ plane. Khan et al. found that the even and odd bands in TMDs have two different band gaps $E_{g\parallel}$ and $E_{g\perp}$, respectively.\cite{Erementchouk2015,Khan2017} $E_{g\perp}$ has been usually neglected for pristine TMDs because of its substantially larger value and weak optical response as compared with $E_{g\parallel}$. Earlier studies\cite{Erementchouk2015,Khan2017,BG_Tune_MoS_2} show that the presence of VDs gives rise to LDS  in addition to the normal extended states present in conduction or valence bands in SL MoS$_{2}$. These LDS appear within the band gap region and they can also be present deep inside the valence band depending on the type of VD. Optical transitions between LDS across Fermi level appear as resonance peaks, both in $\alpha_{\parallel}$  and $\alpha_{\perp}$, which shows that odd states are necessary for understanding the properties of VDs in SL MoS$_{2}$.\cite{Erementchouk2015,Khan2017} The same symmetry considerations apply to Ho:MX$_2$ when Ho substitutes the M atom. 
 
 The goal of this paper is to demonstrate the existence of localized Ho spin-orbit states inside the bandgap of WS$_2$, the ultra-narrow optical transitions due to the atom-like f-orbital states of Ho, and the strict optical selection rules by means of a combination of ab-initio DFT calculations, the Kubo-Greenwood formula, and group theory including SOC.
\begin{figure*}[hbt]
	\begin{center}
		\includegraphics[width=7in]{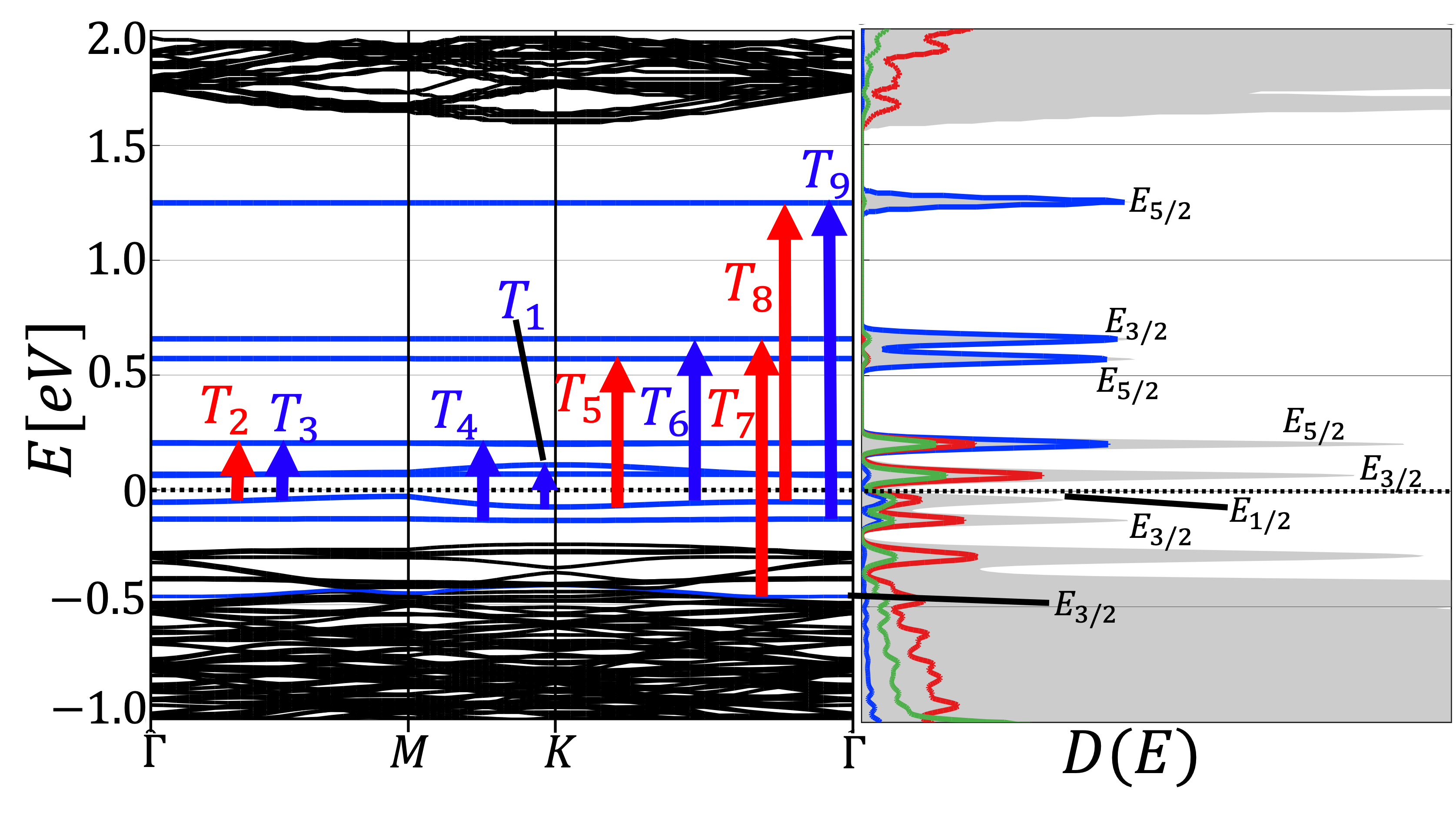}
	\end{center}
	\caption{Bandstructure and density of states, shaded grey region shows the total density of states and the colored curves shows the projected density of states (Blue:$f-$orbitals of the Ho atom, Green:$p-$orbital of the neighboring S atoms and red:$d-$orbitals of the next neighbor W atoms) of $8\times8\times1$ supercell of WS$_2$ containing an Ho$_{\rm W}$ impurity. The LDS are clearly visible as dispersionless (localized) states, some of which lie inside the bandgap, others lie inside the valence band of WS$_2$. The eigenstates corresponding to the LIS transform according to the IRs of the point symmetry group $D_{\rm 3h}$. Vertical arrows indicate optical transitions corresponding to resonances shown in Fig. \ref{fig:OS_Er_W}.}
	\label{fig:bandstructure_Ho_W}
\end{figure*}
 \section{Bandstructure}
 \label{sec:bandstructure}
 
 \subsection{Numerical Analysis}
All Numerical calculations are carried out by using DFT and with the use of \mbox{Perdew–Burke–Ernzerhof} (PBE) generalized gradient (GGA) parametrization\mbox{\cite{PZ_functionals}} for exchange-correlation functional. Fully reletivistic noncollinear and spin polarized DFT calculations were performed as implemented in the Synopsis Atomistix Toolkit (ATK) 2021.06.\mbox{\cite{QW_1}}. For Ho$\rm{_W}$ impurity calculations, we consider a supercell consisting of eight unit cells along each crystalline-axis direction of the monolayer plane (i.e. 64 W atoms and 128 S atoms) and then replace a single W atom with Ho atom as shown in Fig.\mbox{~\ref{fig:Er_W_defect}a)}. We consider a large supercell of edge length 25.22 \AA, to fix the inter-impurity interactions. The point group of WS$_2$ with Ho$_{\rm W}$ defect is D$_{3h}$. The periodic structure of the superlattice allows one to characterize the electron states by the bandstructure $\epsilon_n(\mathbf{k})$, where $\mathbf{k}$ is the vector in the first Brillouin zone of the superlattice and $n$ enumerates different bands.  The sampling of the Brillouin zone was done for a supercell with the equivalent of a 32$\times$32$\times$1 \mbox{Monkhorst–Pack} k-point grid for the WS$_2$ primitive unit cell with a cutoff energy of 400 Ry. For all calculations, structure is first geometrically optimized with a force tolerance of 0.05 eV/\AA. The formation energy for the Ho$_{\rm W}$ impurity is calculated by means of the relation
\begin{equation}
     E^{f}[\rm{Ho_W}]=E_{tot}[\rm{Ho_W}]-E_{tot}[\rm{host}]-\sum_{i}n_{i}\mu_{i}.
\end{equation}
$E_{tot}[{\rm Ho_W}]$ and $E_{tot}[{\rm host}]$ are the total energy of the system with and without the impurity, respectively, $n_i$ is the number of added $(n_i> 0)$ or removed $(n_{i}< 0)$ species of atoms during the formation of the impurity. $\mu_{i}$'s are chemical potentials of the W and Ho atoms, which are estimated from their corresponding bulk forms. The small value of the formation energy $E^{f}[{\rm Ho_W}]=0.846$ eV indicates that the Ho$_{\rm W}$ impurity in  $8\times8\times1$ WS$_2$ is thermally stable.

We first obtain the results for bandstructure and electric susceptibility for pristine WS$\rm{_2}$ as shown in Fig.~\mbox{\ref{fig:Er_W_defect}~b) and c)}, values of the band gap (1.64 eV) and splitting of the valence band edge (425 meV) due to SOC, are in good agreement with previously reported values.\mbox{\cite{SOC_1,SOC_2,SOC_agreement_1,SOC_agreement_2}} The crystal structure of SL WS$_2$ is three atoms thin, where W atom is sandwiched in between two S atoms \mbox{(S-W-S)} via strong covalent bonds. Pristine SL WS$\rm{_2}$ is invariant with respect to $\sigma_h$ reflection about the z = 0 (W) plane, where the z$-$axis is oriented perpendicular to the W plane of atoms. Therefore, electron states break down into two classes: even and odd, or symmetric and antisymmetric with respect to $\sigma_h$. $d$-orbitals of the W and $p^{(t,b)}$- orbitals ($t$ and $b$ denoting the top and bottom layers) of the S atoms  give the largest contribution to the conduction and valence band structure of SL WS$_2$. \mbox{\cite{SOC_agreement_1,Guinea_tight_binding_model}} Based on the \mbox{$\sigma_{h}$} symmetry, the even and odd atomic orbitals are spanned by the bases $\{\phi_1=d_{x^2 - y^2}^W, {~}\phi_2=d_{xy}^W, {~}\phi_3=d_{z^2}^W,{~} \phi_{4,5}={~}p_{x,y}^{e}=(p_{x,y}^{(t)} + p_{x,y}^{(b)})/\sqrt{2},  {~}\phi_6={~}p_{z}^{e}=(p_{z}^{(t)} - p_{z}^{(b)})/\sqrt{2} \}$  and $\{\phi_7=d_{xz}^W, {~}\phi_8=d_{yz}^W, {~}\phi_{9,10}=p_{x,y}^{o}=(p_{x,y}^{(t)} - p_{x,y}^{(b)})/\sqrt{2}, {~}\phi_{11}=p_{z}^{o}=(p_{z}^{(t)} + p_{z}^{(b)})/\sqrt{2}\}$, respectively.

 \begin{figure*}[hbt]
	\begin{center}
		\includegraphics[width=7in]{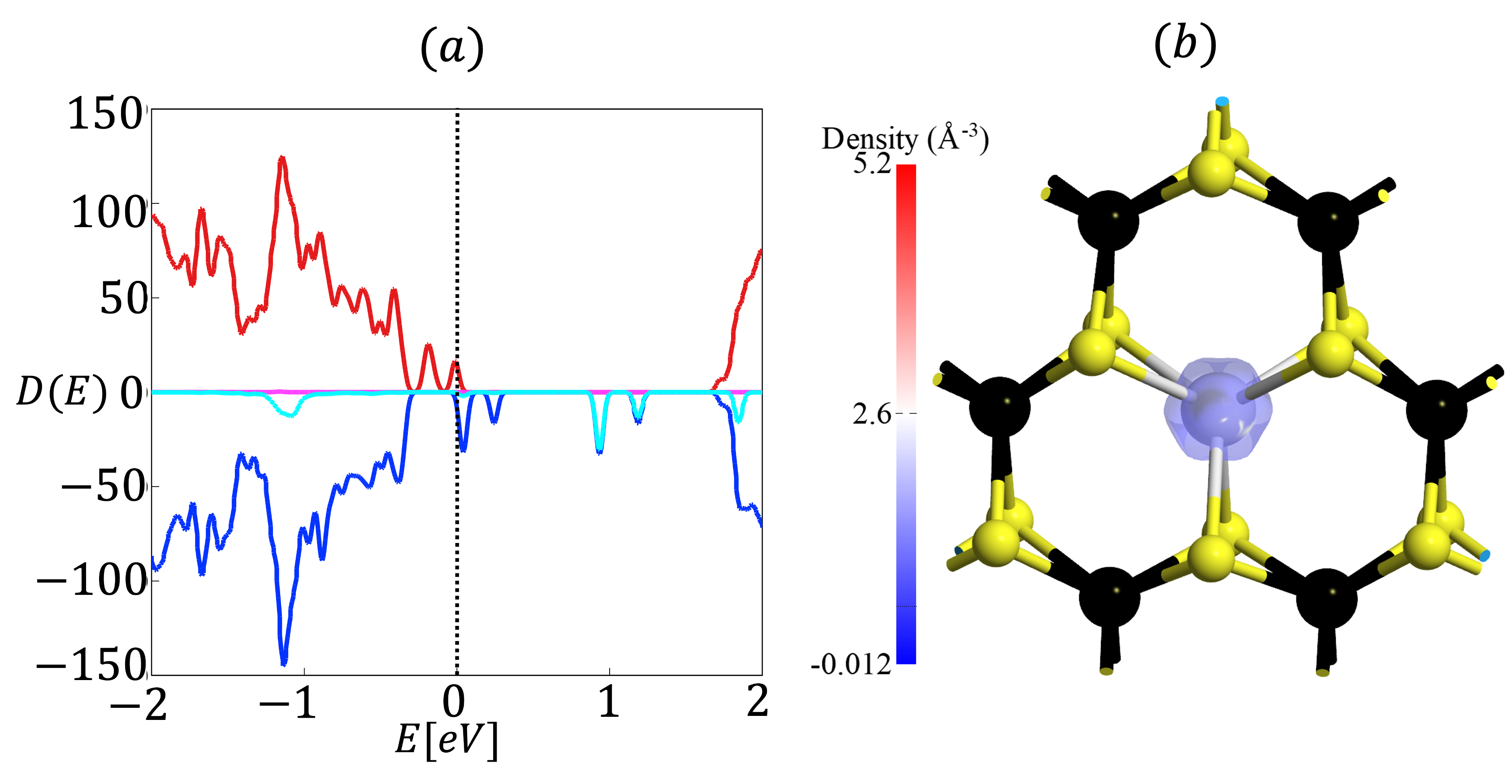}
	\end{center}
	\caption{a) Spin-polarized density of states for Ho$\rm_W$:SL WS$_2$. Red(blue) is for spin-up(down) total Density of states while cyan(magenta) is for spin-up(down) projected density of states for Ho atom. b) spin density $\rho_{\uparrow}-\rho_{\downarrow}$, concentrated on the Ho atom with a magnetic moment of 4.75$\mu_{B}$. Spin density is plotted for an isovalue of 0.08911 \AA$^{-3}$.}
	\label{fig:mag_sig}
\end{figure*}
 
\begin{figure*}[htb]
\centering
\includegraphics[width=7in]{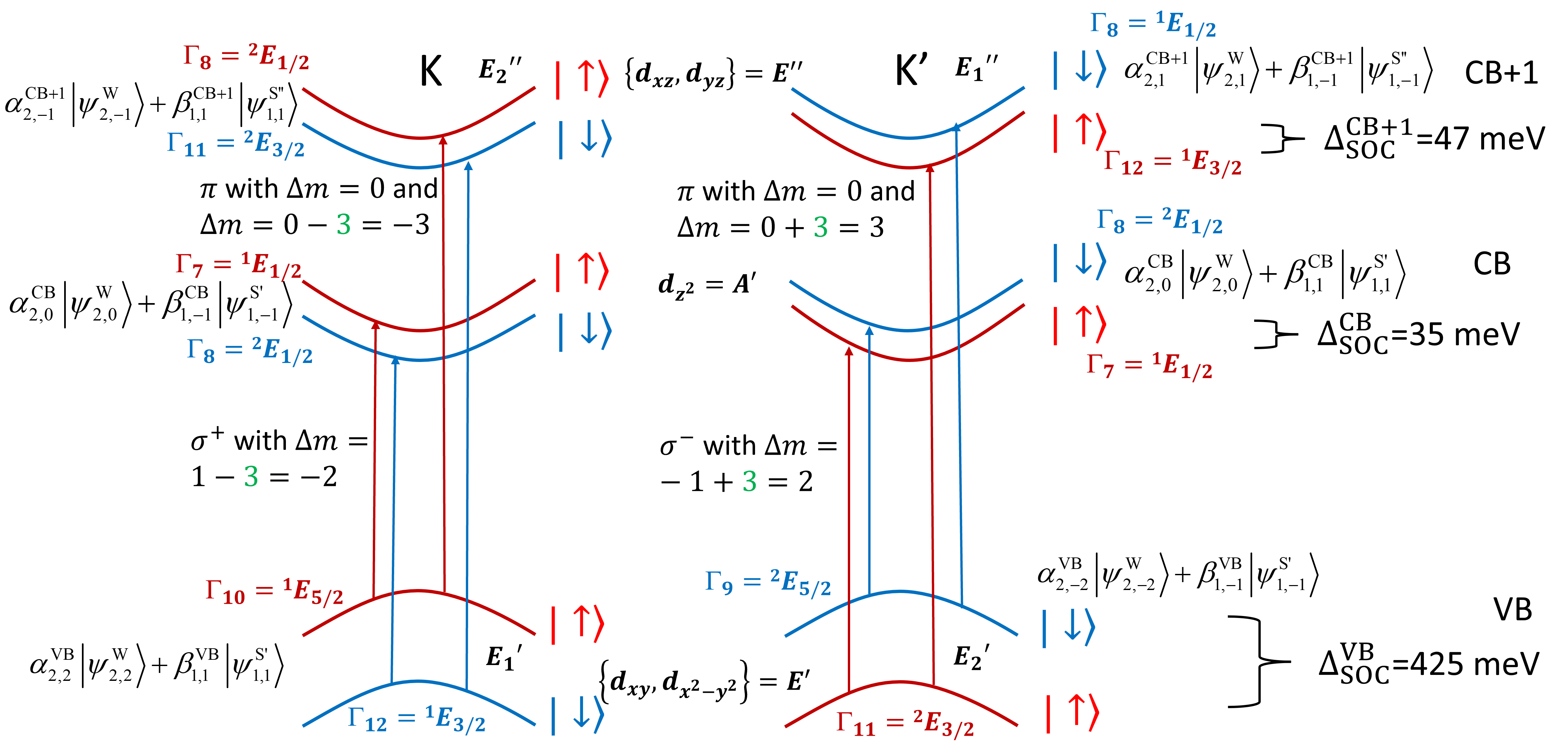}
\caption{The optical selection rules in pristine SL WS$_2$ satisfy the equation $\Delta m=\pm 1\pm 3$ for $\sigma^\pm$ transitions and $\Delta m=0\pm 3$ for $\pi$ transitions. The term $\pm 3$ is due to the $C_3$ rotational symmetry of the lattice. These selection rules corroborate the difference between the in-plane and out-of-plane band gaps $E_{g||}$ and $E_{g\perp}$, respectively. The Bloch states at the K and K' points transform according to the IRs of the $C_{3h}$ point group. The coefficients of the main contributions are given by $\left|\alpha_{2,\pm 2}^{\rm VB}\right|^2=0.75$, $\left|\beta_{1,\pm 1}^{\rm VB}\right|^2=0.25$, $\left|\alpha_{2,0}^{\rm CB}\right|^2=0.75$, $\left|\beta_{1,\pm 1}^{\rm CB}\right|^2=0.19$, $\left|\alpha_{2,\pm 1}^{\rm CB+1}\right|^2=0.56$, $\left|\beta_{1,\pm 1}^{\rm CB+1}\right|^2=0.29$.}
\label{fig:optical_selection_rules}
\end{figure*}

 Using first principle studies\mbox{\cite{SOC_1,SOC_2,SOC_agreement_1}}, it is known that valence and conduction bands are primarily made from $d_{x^2-y^2}$, $d_{xy}$ and $d_{z^2}$ of W atoms, which transform as E$^{\prime}_1$, E$^{\prime}_2$ and A$^\prime$ irreducible representations (IRs) of the C$_{3h}$ symmetry group at the K and K' points, in the absence of SOC \mbox{(Table~\ref{table_D_3h})}. The presence of SOC couples the spin and orbital angular momenta, thereby requiring the consideration of the double-group IRs. Double-group IRs can be obtained by multiplying single-group IRs with $E_{1/2}$ as shown in Table~\mbox{\ref{table_C_3h}}, where $E_{1/2}$ is the 2D spin representation.
 The spin-orbit states for pristine SL WS$_2$ are shown in Fig.~\ref{fig:optical_selection_rules}.

\begin{table}[h]%
\centering%
\begin{tabular}{|c |c || c |c |c |c |c |c || c | c |}\hline
\multicolumn{2}{|c||}{$C_{3h}$}  &  $E$ &  $C_{3}$   &  $C_{3}^2$     &     $\sigma_{h}$    &    $S_{3}$    &    $S_{3}^5$ & linear & quadratic \\ 
\hline
$A'$  & $\Gamma_1$ &  1  & 1  & 1   & 1   &  1 &  1 & $R_z$ & $x^2+y^2$, $z^2$  \\
$A''$  & $\Gamma_4$ &  1  & 1   & 1  &  $-$1   &    $-$1   & $-$1 & $z$ &  \\
$E'_{1}$ & $\Gamma_2$ &  1 & $\xi$  & $\xi^2$  &  1  & $\xi$  & $\xi^2$ & $x+iy$ & \multirow{2}{*}{$(x^2-y^2,xy)$}    \\
$E'_{2}$ & $\Gamma_3$ &  1 & $\xi^2$  & $\xi$ & 1  & $\xi^2$ &  $\xi$ & $x-iy$ &  \\
$E''_{1}$ & $\Gamma_5$ &  1 & $\xi$ & $\xi^2$  & $-$1 &  $-\xi$ &  $-\xi^2$ & $R_x+iR_y$ & \multirow{2}{*}{$(xz,yz)$}  \\
$E''_{2}$ & $\Gamma_6$ &  1 & $\xi^2$ & $\xi$ & $-$1  & $-\xi^2$ &  $-\xi$ & $R_x-iR_y$ &  \\
\hline
$^1E_{1/2}$ & $\Gamma_7$ &  1 & $-\xi^2$ & $-\xi$ & $i$ &  $i\xi^2$   &  $-i\xi$ & \multicolumn{2}{c|}{$\left|\frac{1}{2},\frac{1}{2}\right>,\left|\frac{3}{2},\frac{1}{2}\right>$} \\
$^2E_{1/2}$ & $\Gamma_8$ &  1 & $-\xi$ & $-\xi^2$ & $-i$ &  $-i\xi$   &  $i\xi^2$ & \multicolumn{2}{c|}{$\left|\frac{1}{2},-\frac{1}{2}\right>,\left|\frac{3}{2},-\frac{1}{2}\right>$} \\
$^1E_{3/2}$ & $\Gamma_{11}$ &  1 & $-1$ & $-1$ & $i$ &  $i$   &  $-i$ & \multicolumn{2}{c|}{$\left|\frac{3}{2},\frac{3}{2}\right>,\left|\frac{5}{2},\frac{3}{2}\right>$} \\
$^2E_{3/2}$ & $\Gamma_{12}$ &  1 & $-1$ & $-1$ & $-i$ &  $-i$   &  $i$ & \multicolumn{2}{c|}{$\left|\frac{3}{2},-\frac{3}{2}\right>,\left|\frac{5}{2},-\frac{3}{2}\right>$}  \\
$^1E_{5/2}$ & $\Gamma_9$ &  1 & $-\xi$ & $-\xi^2$ & $i$ &  $i\xi$   &  $-i\xi^2$ & \multicolumn{2}{c|}{$\left|\frac{5}{2},\frac{5}{2}\right>$} \\
$^2E_{5/2}$ & $\Gamma_{10}$ &  1 & $-\xi^2$ & $-\xi$ & $-i$ &  $-i\xi^2$   &  $i\xi$ & \multicolumn{2}{c|}{$\left|\frac{5}{2},-\frac{5}{2}\right>$} \\
\hline\end{tabular}
\caption{Character table of the group $C_{3h}$ with $\xi^3=1$. Two common notations are used for the IRs of the single and double group. The reduction of symmetry from $D_{3h}$ to $C_{3h}$ is accompanied by the compatibility relations $A_1',A_2'\rightarrow A'$, $E'\rightarrow E_1'\bigoplus E_2'$, $A_1'',A_2''\rightarrow A''$, $E''\rightarrow E_1''\bigoplus E_2''$, $E_{1/2}\rightarrow {^1}E_{1/2}\bigoplus {^2}E_{1/2}$, $E_{3/2}\rightarrow {^1}E_{3/2}\bigoplus {^2}E_{3/2}$, and $E_{5/2}\rightarrow {^1}E_{5/2}\bigoplus {^2}E_{5/2}$.}
\label{table_C_3h}
\end{table}
 
 The bandstructure of WS$_2$ with Ho$_{\rm W}$ impurities is shown in Fig.~\ref{fig:bandstructure_Ho_W}. Regular electronic states within the valence or conduction bands are depicted by black lines while LIS ($f-$orbitals of Ho) are depicted by blue lines. Some of the allowed optical transitions between different $f-$orbitals of Ho are depicted by vertical arrows. The resulting optical spectrum is shown in Fig.~\ref{fig:OS_Er_W}.

 The Kramers theorem states that for every energy eigenstate of a time-reversal symmetric system with half-integer total spin, there is at least one more eigenstate with the same energy. In other words, every energy level is at least doubly degenerate if it has half-integer spin. It can be seen that Kramers degeneracy, which is a consequence of time reversal symmetry, is broken for LIS in Ho$_{\rm W}$:SL WS$_2$.  In Ref. \onlinecite{Ho_Mo_MS} it has been shown that presence of Ho$_{\rm Mo}$ impurity leads to spin polarization and results in long range ferromagnetic coupling between local spins. The local magnetic moment of the Ho$_{\rm W}$ impurity breaks the time reversal symmetry and lifts the Kramers degeneracy. In order to confirm that indeed the Ho$_{\rm W}$ impurity in SL WS$_2$ contains a magnetic moment, DFT calculations are performed by using spin-polarized GGA method. The results are presented in Fig.~\ref{fig:mag_sig}, where we show that the Ho impurity has a magnetic moment of 4.75$\mu_B$.  Our spin-polarized DFT calculations show that the exchange correlation potential leads to a spin splitting for Ho$_{\rm W}$:SL WS$_2$. In Fig.~\ref{fig:mag_sig} (b) the isosurface plot for spin density shows that the main contribution to the magnetism is due to the $f-$orbitals of Ho atom while the bulk states do not show any magnetic moment, in contrast to what has been observed in Ref.~\onlinecite{Ho_Mo_MS}, where a long-range magnetic interaction is seen. The reason is that we consider a much larger supercell of $8 \times 8 \times 1$, as opposed to their $4 \times 4 \times 1$ supercell, resulting in a dilution of the impurity concentration that suppresses long-range magnetic interaction.   %Magnetic properties of rare earth doped TMDCs is an The details of magnetic properties of Ho$\rm_W$:SL WS$_2$ will be presented somewhere else.
 
 \begin{table}[h]%
\centering%
\begin{tabular}{ |c | c ||c |c |c |c |c |c || c | c|}\hline
\multicolumn{2}{|c||}{$D_{3h}$}   &    $E$    &    $\sigma_{h}$    &    $2C_{3}$    &    $2S_{3}$    &    $3C_{2}$    &    $3\sigma_{v}$ & linear & quadratic\\ 
\hline
$A'_{1}$ & $\Gamma_1$ & 1 & 1   & 1  & 1   &   1   &  1 & &  $x^2+y^2$, $z^2$  \\
$A'_{2}$ & $\Gamma_2$ & 1 & 1  & 1 &  1  & $-$1 & $-$1 & $R_z$ &  \\
$A''_{1}$ & $\Gamma_3$ & 1 & $-$1  & 1  &  $-$1    &  1 & $-$1 & &  \\
$A''_{2}$ & $\Gamma_4$ & 1 & $-$1 & 1 & $-$1  &  $-$1 &  1  & z &  \\
$E'$ & $\Gamma_6$ & 2 & 2 & $-$1 & $-$1  & 0 &  0 & $(x,y)$ & $(x^2-y^2,xy)$ \\
$E''$ & $\Gamma_5$ & 2   & $-$2  & $-$1  & 1 &  0  &  0 & $(R_x,R_y)$ & $(xz,yz)$\\
\hline
$E_{1/2}$ & $\Gamma_7$  & $\pm 2$ & 0 & $\pm 1$  & $\pm\sqrt{3}$ &  0 &  0 & \multicolumn{2}{c|}{$\left|\frac{1}{2},\pm\frac{1}{2}\right>,\left|\frac{3}{2},\pm\frac{1}{2}\right>$} \\
$E_{3/2}$ & $\Gamma_9$ & $\pm 2$ & 0 & $\mp 2$ & 0  &  0  &  0  & \multicolumn{2}{c|}{$\left|\frac{3}{2},\pm\frac{3}{2}\right>,\left|\frac{5}{2},\pm\frac{3}{2}\right>$} \\
$E_{5/2}$ & $\Gamma_8$ & $\pm 2$ & 0 & $\pm 1$ & $\mp\sqrt{3}$ & 0 &  0 & \multicolumn{2}{c|}{$\left|\frac{5}{2},\pm\frac{5}{2}\right>$} \\
\hline\end{tabular}
\caption{Character table of the group $D_{3h}$. Two common notations are used for the IRs of the single and double group.}
\label{table_D_3h}
\end{table}

\begin{table}[h]%
\centering%
\begin{tabular}{ |c |c |c |c |c |c |c |}\hline
 $\Gamma_{i}(D_{3h})$                          &    $A^{\prime}_{1}$    &    $A^{\prime}_{2}$    &    $A^{\prime\prime}_{1}$    &    $A^{\prime\prime}_{2}$    &    $E^{\prime}$    &    $E^{\prime\prime}$\\ 
\hline
$\Gamma_{i}\bigotimes E_{1/2}$   & $E_{1/2}$ & $E_{1/2}$  & $E_{5/2}$  & $E_{5/2}$   &  $E_{3/2}\bigoplus E_{5/2}$  &  $E_{1/2}\bigoplus E_{3/2}$\\
\hline 
\end{tabular}
\caption{Double-group representations obtained from single-group representation for $D_{3h}$ group.}
\label{table_D_3h_DG}
\end{table}
 
 %The Ho$_W$ defect preserves the $\sigma_{h}$ symmetry of the crystal and can be described by the group $D_{3h}$.\cite{Cheiwchammangij_SOC,Song_SOI} Symmetry operations and different irreducible representations (IRs) of D$_{3h}$ symmetry are shown in Table~\ref{table_D_3h}.  Due to $\sigma_h$ symmetry of Er$_W$ defect, the LIS break down into even and odd parity with respect to the W-plane of atoms. It can be seen in Fig.~\ref{fig:bandstructure_Er_W} and \ref{fig:Bloch_states} that LIS appear within the bandstructure as even (blue) and odd (red) states. In addition, the LIS appear in the form of triplets, i.e. a degenerate doublet and a singlet, which is in fact a consequence of 3-fold rotation symmetry $C_3$ of the defect,\cite{Erementchouk2015,Khan2017} as shown in Fig.~\ref{fig:Bloch_states}. 
 
\begin{figure*}[hbt]
\centering
\includegraphics[width=7in]{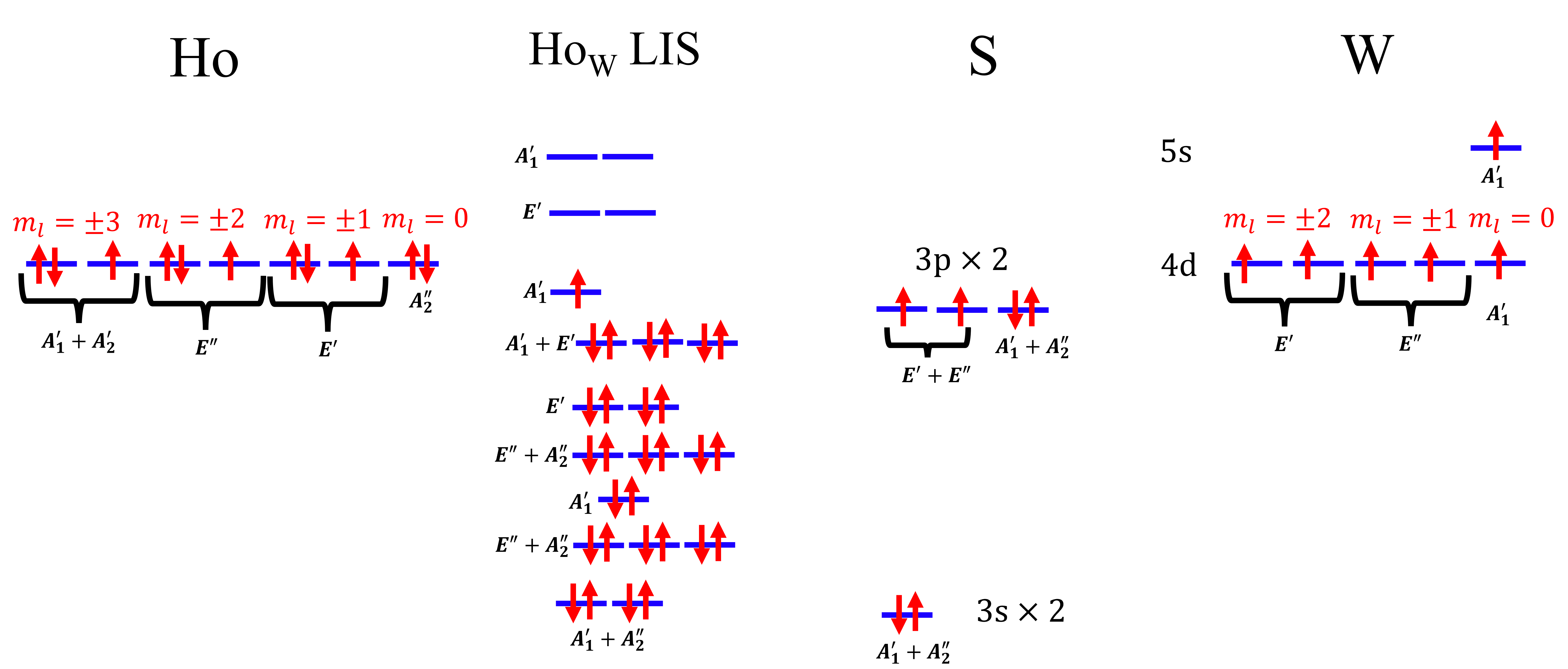}
\caption{Molecular orbital diagram of Ho f orbitals in WS$_2$, giving rise to the Ho$_W$ LIS shown in the bandstructure in Fig.~\ref{fig:bandstructure_Ho_W}. The states are labeled with the IRs of the point group $D_{3h}$.}
\label{fig:MOT_HoW}
\end{figure*}

 \subsection{Molecular Orbital Theory}
A Ho$\rm_W$ impurity inside WS$_2$ looks similar to an atom in an effective electrostatic ligand field created by its neighboring six sulphur atoms. In this approximation molecular orbital theory (MOT) can be used. To identify the LIS in the DOS, the projected density of states (PDOS) showing orbital contributions of individual atoms is shown in Fig.~\ref{fig:bandstructure_Ho_W}. In addition to the contribution from the $f$ orbitals of Ho$_{\rm W}$, contributions from the $p$ orbitals of the nearest neighboring S atoms and from the $d$ orbitals of next-nearest neighboring W atoms are present. This means that in the Hilbert base spanned by $\psi_{i}^\dagger=(\phi_{1},...,{~}\phi_{11},
 %\phi_{12}=d_{x^2 - y^2}^{Er}, {~}\phi_{13}=d_{xy}^{Er}, {~}\phi_{14}=d_{z^2}^{Er},{~}\phi_{15}=d_{xz}^{Er},{~}\phi_{12}=d_{yz}^{Er},
 {~}\phi_{12}=f_{z^3},{~}\phi_{13}=f_{xz^2},{~}\phi_{14}=f_{yz^2},{~}\phi_{15}=f_{xyz},{~}\phi_{16}=f_{z(x^2-y^2)},{~}\phi_{17}=f_{x(x^2-3y^2)},{~}\phi_{18}=f_{y(3x^2-y^2)})^{\dagger}$, an LIS state can be represented by 
 \begin{equation}
     \Psi=\sum_{j}a_{j}\phi_{j},
 \end{equation} 
where the real coefficients $a_{i}$'s can be extracted from the PDOS shown in Fig.~\ref{fig:bandstructure_Ho_W}. Since admixture of orbitals is only allowed if they belong to the same IR, many coefficients are zero. 
The MOT diagram of pristine WS$_2$ can be found in  Ref.~\onlinecite{ER_W_khan}. The resulting eigenstates, identified by their IRs of  $D_{3h}$, match the continuum states of the bands in WS$_2$, as can be seen from Ref.~\onlinecite{Pike2017}.

Analyzing the PDOS, it becomes obvious that the Ho $f$ orbitals couple to both the $p$ orbitals of nearest neighboring S atoms and $d$ orbitals of next-nearest neighboring W atoms. The resulting MOT diagram including Ho LIS is shown in Fig.~\ref{fig:MOT_HoW}.
The orbital energy ordering can be determined by comparison with the PDOS shown in Fig.~\ref{fig:bandstructure_Ho_W}.
The highest occupied molecular orbital (HOMO) is a  $E_{1/2}$ spin-orbit state  with an orbital $A_1'$ singlet state. The lowest unoccupied molecular orbital (LUMO) is a $E_{3/2}$ spin-orbit state with a $E'$ orbital doublet state, which matches the PDOS in Fig.~\ref{fig:bandstructure_Ho_W}.
Although the Ho atom with an average atomic radius of 1.75 {\AA} is substantially larger than a W atom with an average atomic radius of 1.35 \AA, DFT shows that the Ho$_{\rm W}$ impurity is stable in the WS$_2$ host crystal.
Because of the strong lattice distortions there are relatively strong hybridizations between the Ho $f$ orbitals and the W $d$ orbitals, as can be seen in the bandstructure in Fig.~\ref{fig:bandstructure_Ho_W}.

\section{Optical Response}
\label{sec:optical}

\begin{figure} %[optical_response]
	\begin{center}
		\includegraphics[width=3.5in]{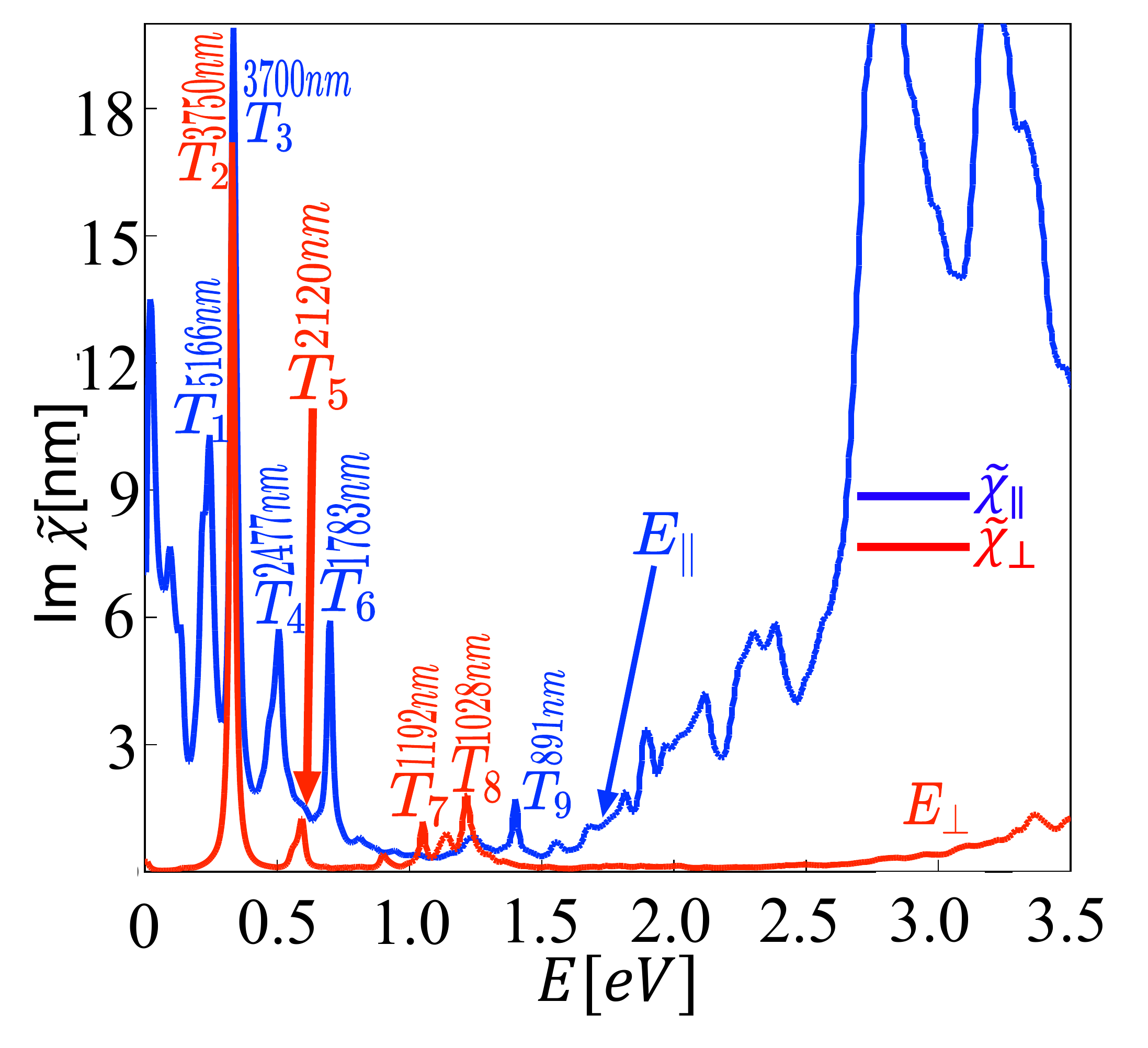}
	\end{center}
	\caption{Optical spectrum calculated by means of ATK showing resonances of Im$[\epsilon_{\parallel}](\omega)$ (blue) and Im$[\epsilon_{\perp}](\omega)$ (red) $a)$ due to Ho$_{\rm W}$ impurities in WS$_2$.}
	\label{fig:OS_Er_W}
\end{figure}

Since the f-orbital contribution to the LIS is large, the optical spectrum exhibits narrow peaks, reminding of atom-like optical transitions. The relative dielectric functions $\epsilon_r$ of various TMDs have been measured in Ref.~\onlinecite{Dielectric_function_measurement}. We evaluate the matrix elements of the dielectric tensor in three dimensions ($i,j=x,y,z$) using the Kubo–Greenwood formula for the electric susceptibility
\begin{equation}\label{eq:KGW}
\chi_{ij}(\omega)=\frac{e^{2}}{\hbar m_{e}^{2}V}\sum_{uv\bf{k}}\frac{f_{u\bf{k}}-f_{v\bf{k}}}{\omega_{uv}^2({\bf{k}})[\omega_{uv}({\bf{k}})- \omega-i\Gamma/{\hbar}]}p_{uv}^{i}p_{vu}^{j}
\end{equation}  
where $p_{pq}^{j}=\langle u{\bf{k}}|p^{j}|v\bf{k}\rangle$ is the dipole matrix element between Bloch states $\langle\bf{r}$$|u$$\bf{k}\rangle= \psi_{{u}\mathbf{k}}(\bf{r})$ and $\langle\bf{r}$$|v$$\bf{k}\rangle= \psi_{{v}\mathbf{k}}(\bf{r})$, $V$ the volume of the crystal, $f$ the Fermi function, and $\Gamma=0.01$ eV the broadening. A vacuum separation of $a_{3}=20$ {\AA} has been chosen in order to suppress not only electron bonding but also electrostatic interactions.  In this limit the Bloch functions are localized on SL WS$_2$. Consequently, we can use the approximation $(1/V)\sum_{k_z}\rightarrow(1/\Omega a_{3})$, where $\Omega$ is the surface area of SL WS$_2$. In this case $\tilde{\chi}=a_{3}\chi$, which has the unit of length, is independent of the vacuum separation.  
Using this definition, we present the in plane $\chi_{\parallel}$ and out of plane $\chi_{\perp}$ components of the 3D susceptibility tensor for Ho$_{\rm W}$ impurities in SL WS$_2$ in Fig.~\ref{fig:OS_Er_W}. 
 We focus on transitions between states near the conduction and valence band edges and inside the band gap with resonance frequency $\hbar\omega_{uv}=|\varepsilon_{u}-\varepsilon_{v}|$, where $\varepsilon_u$ is the eigenenergy of the Bloch state $\psi_{u\bf{k}}(\bf{r})$. 
 
 For pristine SL WS$_2$ the point group symmetry at the K and K' points is $C_{3h}$. A general result from group theory states that an optical transition is allowed by symmetry only if the direct product $\Gamma(|v\bf{k}\rangle)\otimes$ $\Gamma(p^j)$$\otimes$$\Gamma(|u\bf{k}\rangle)$ contains $\Gamma(I)$ in its decomposition in terms of a direct sum. $\Gamma(I)$ denotes the IR for the identity, i.e., $A'$ for $C_{3h}$.
 The in plane and out of plane components of $p_{vu}^{j}$ must be considered individually because they transform according to different IRs of the point group. The resulting optical selection rules are shown in Fig.~\ref{fig:optical_selection_rules}. These selection rules corroborate the difference between the in-plane and out-of-plane band gaps $E_{g||}$ and $E_{g\perp}$, respectively, which can be seen in the in-plane and out-of-plane susceptibilities Im$[\chi_{\parallel}](\omega)$ and Im$[\chi_{\perp}](\omega)$, respectively [see Fig.~\ref{fig:Er_W_defect}]. We predicted this difference in Refs.~\onlinecite{Erementchouk2015,Khan2017}, which has later been experimentally confirmed.\cite{Wang2017} This difference has also been verified theoretically by means of DFT calculations with GW correction and the solution of the Bethe-Salpeter equation for in-plane and out-of-plane excitons.\cite{Guilhon2019}
 
 \begin{table}[h]
\begin{tabular}{|c |c |c |c |c |c |c|}\hline
$C_{3h}$ & $A^{\prime}$ & $A^{\prime\prime}$ & $E_{1}^{\prime}$ & $E_{2}^{\prime}$ & $E_1^{\prime\prime}$ & $E_2^{\prime\prime}$   \\
\hline
$A^{\prime}$ & & $\pi$ &  $\sigma^-$  & $\sigma^+$      &    &   \\
\hline
 $A^{\prime\prime}$  & $\pi$ &  &   &  & $\sigma^-$ &  $\sigma^+$   \\
\hline
$E_{1}^{\prime}$  &  $\sigma^-$ & & $\sigma^+$ & &   & $\pi$      \\
\hline
$E_{2}^{\prime}$ &  $\sigma^+$ & & & $\sigma^-$ & $\pi$ & \\
\hline
$E_1^{\prime\prime}$ &  & $\sigma^-$ & & $\pi$ & $\sigma^+$ & \\
\hline
$E_2^{\prime\prime}$ & & $\sigma^+$ & $\pi$ & & & $\sigma^-$ \\
\hline
\end{tabular}
\caption{Electric Dipole selection rules in $C_{3h}$ symmetry. $\sigma$ represents in plane transitions while $\pi$ represents out of plane transitions.}
\label{table_D_3h_C_3v}
\end{table}

Alternatively, it is possible to use the conservation of angular momentum to derive the optical selection rules. In pristine SL WS$_2$ the $C_3$ rotational symmetry relaxes the atomic optical selection rules $\Delta m=\pm 1$ for $\sigma^\pm$ transitions and $\Delta m=0$ for $\pi$ transitions to $\Delta m=\pm 1\pm 3$ for $\sigma^\pm$ transitions and $\Delta m=0\pm 3$ for $\pi$ transitions, whereby an angular momentum mismatch of $\pm 3$ can be transferred to or from the crystal lattice.
The resulting optical selection rules match the ones obtained above from group theory and are also shown in Fig.~\ref{fig:optical_selection_rules}.
When using the approximation of a two-band model described by a Dirac Hamiltonian for the conduction (CB) and valence (VB) bands, our selection rules match the ones shown in Ref.~\onlinecite{Xiao2015}.

 \begin{figure*}
\centering
\includegraphics[width=7.2in]{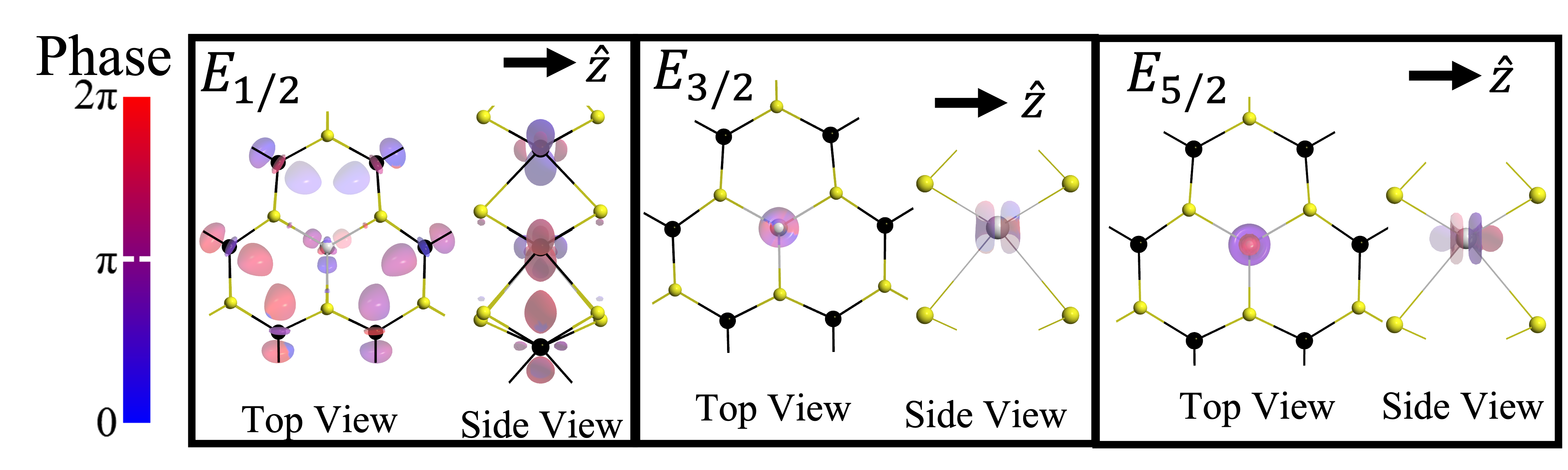}
\caption{Examples of the Bloch states for the Ho$_W$ impurity in $8\times8\times1$ super cell of WS$_2$.}
\label{fig:Bloch_states}
\end{figure*}

Given the point group symmetry of impurities in a crystal, the LIS transform according to its IRs. In the case of the Ho$_{\rm W}$ impurity the point group symmetry is $D_{3h}$, its character table shown in Table~\ref{table_D_3h}. 
The identity for $D_{3h}$ is $A_{1}'$. Table~\ref{table_D_3h_C_3v} shows the the selection rules for electric dipole transitions for the IRs. Note that the electromagnetic field couples only to the orbital part of the Bloch states. Therefore the we need to consider only the orbital IRs of $D_{3h}$. Remarkably, we show in Table \ref{tab:comparison} that several optical transitions are in good agreement with available experimental data for optical transitions in Ho$^{3+}$:YAG.  
%We find that the out-of-plane component of ${\bf{P}}_{pq}^{j}$, which gives rise to $\pi$ transitions, is nonzero only between odd and even states of the same multiplicity (either between singlets or between doublets), while the in-plane components, which lead to σ transitions, are nonzero for all states of the same parity.The optical transitions are in agreement with the selection rules derived in Table~\ref{table_D_3h_C_3v}.

\begin{table}[h]
\begin{tabular}{|c |c |c |c |c |c |c|}\hline
$D_{3h}$ & $A_{1}^{\prime}$ & $A_{2}^{\prime}$ & $A_{1}^{\prime\prime}$ & $A_{2}^{\prime\prime}$ & $E^{\prime}$ & $E^{\prime\prime}$   \\
\hline
$A_{1}^{\prime}$ & & &    & $\pi$      &  $\sigma$  &   \\
\hline
 $A_{2}^{\prime}$  &  &     & $\pi$   &               & $\sigma$ &     \\
\hline
$A_{1}^{\prime\prime}$  &      & $\pi $  & & &     & $\sigma$      \\
\hline
$A_{2}^{\prime\prime}$ &  $\pi$ & & & & & $\sigma$\\
\hline
$E^{\prime}$ & $\sigma$&$\sigma$ & & & $\sigma$&$\pi$\\
\hline
$E^{\prime\prime}$ & & & $\sigma$&$\sigma$ & $\pi$&$\sigma$\\
\hline
\end{tabular}
\caption{Electric Dipole selection rules in $D_{3h}$ symmetry. $\sigma$ represents in plane transitions while $\pi$ represents out of plane transitions.}
\label{table_D_3h_C_3v}
\end{table}

\begin{table}[h]
\begin{tabular}{|c ||c |c |}\hline
Ho$^{3+}$ states & Ho$^{3+}$:YAG & Ho$^{3+}$:WS$_2$   \\
\hline\hline
$^5$I$_{4}\rightarrow$ $^5$I$_{5}$  &  &  5.1 $\mu$m ($T_1$) \\
\hline
$^5$I$_{5}\rightarrow$ $^5$I$_{6}$  &  &  3.75 $\mu$m ($T_2$) or 3.70 $\mu$m ($T_3$) \\
\hline
$^5$I$_{6}\rightarrow$ $^5$I$_{7}$  &  &  2.47 $\mu$m ($T_4$)  \\
\hline
$^5$I$_{7}\rightarrow$ $^5$I$_{8}$  & \begin{tabular}{c} 1.978 $\mu$m [\onlinecite{Malinowski2000}] \\ 2.09 $\mu$m [\onlinecite{Ho_YAG}] \end{tabular} &  2.12 $\mu$m ($T_5$) or 1.78 $\mu$m ($T_6$) \\
\hline
$^5$I$_{6}\rightarrow$ $^5$I$_{8}$  & 1.169 $\mu$m [\onlinecite{Malinowski2000}] &  1.19 $\mu$m ($T_7$) or 1.10 $\mu$m ($T_8$)  \\
\hline
$^5$I$_{5}\rightarrow$ $^5$I$_{8}$  & 899 nm [\onlinecite{Malinowski2000}] &  891 nm ($T_9$)  \\
\hline
\end{tabular}
\caption{Resonance wavelength of optical transitions (absorption and emission) for Ho$^{3+}$:YAG and Ho$^{3+}$:WS$_2$. The absorption transitions in Ho$^{3+}$:WS$_2$ are labeled according the optical spectrum shown in Fig.~\ref{fig:OS_Er_W}.}
\label{tab:comparison}
\end{table}

\section{Conclusion}
Our results of electronic and optical properties of Ho$_{\rm W}$ impurities in SL WS$_2$ reveal LIS inside and near the band gap and atom-like sharp optical transitions both in $\chi_\parallel$ and $\chi_\perp$. Therefore, we argue that REAs in TMDs are good candidate for spin qubits. Let us elaborate further. 

%During the preparation of this manuscript we learned of similar ab-initio calculations for electronic and optical properties of Er$_{\rm W}$ defects in WS$_2$ using VASP.\cite{Lopez2021}While VASP is based on plane-wave basis sets, ATK is based on atomic orbital basis sets. That is why there are differences in the results.

Atom-like sharp optical transitions suggest that the decoherence time should be very long, which is one of the main criteria for a spin qubit. As mentioned in the introduction, by choosing a host material free  of paramagnetic impurities and nuclear spins, it would be possible to substantially increase the spin coherence time of the impurity spin, in this case the spin of a Ho$_{\rm W}$ impurity.
Therefore, let us compare rare-earth atom spins in TMDs with other currently existing solid-state spin qubits: 
\begin{itemize}
\item W has a weak abundance of 14\% of nuclear spin ½ ($^{183}$W) and S has a negligibly small abundance of 0.8\% of nuclear spin 3/2 ($^{33}$S). These can be removed by isotopic purification. In stark contrast to that, electron spin qubits in quantum dots made of GaAs suffer from hyperfine interaction. The issue is that Ga and As cannot be isotopically purified because all naturally abundant Ga and As isotopes have nuclear spins. Comparing to NV centers in diamond, N has a 99.6\% abundance of nuclear spin 1 ($^{14}$N) and 0.4\% of nuclear spin 1/2 ($^{15}$N). Therefore the nuclear spins of nitrogen cannot be removed by isotopic purification, either. In addition, P1 N impurities and surface spins are paramagnetic impurities that also lead to decoherence of an NV qubit. Consequently, we expect a much weaker decoherence of the Ho spin state.
\item The location in the direction perpendicular to the plane of the 2D material of the rare-earth impurities is accurate on the atomic level. By contrast, in 3D materials, such as GaAs and diamond, impurities and defects are spread throughout the 3D materials. Therefore we expect enhanced quantum sensing due to accurate distance to target atoms.
\item 2D materials have clean surfaces, in stark contrast to diamond that hosts dark P1 nitrogen impurities with nuclear spins and surface spins. The spin coherence time of shallow NV centers in diamond within 30 nm of the surface degrades drastically due to increased electric and magnetic noise. Diamond surfaces are difficult to be etched and polished in a controlled way due to diamond's hardness. \cite{Leon2021}
\end{itemize}
Ce$^{3+}$ in YAG exhibits electronic decoherence times of $T_2=2$ ms under dynamic decoupling \cite{Siyushev2014}. This is relatively long considering that $^{27}$Al, the only naturally occurring isotope, has a nuclear spin of 5/2. Liu et al. have recently performed the Deutsch-Jozsa quantum algorithm on the electron spin of a Ce$^{3+}$ ion in YAG by means of phase gates with an operation time of $t_{op}=0.3$ $\mu$s.\cite{Liu2020} This would allow for $N=T_2/t_{op}=6.66\times 10^3$ quantum operations.
In the case of Er$^{3+}$ impurities in CaWO$_4$, the spin coherence time is $T_2=23$ ms, without isotopic purification.\cite{Dantec2021} In principle, this would allow for $N=T_2/t_{op}=6.66\times 10^4$ quantum operations.

Since WS$_2$ can be isotopically purified to have zero nuclear spin, we expect even longer decoherence times and better performance with REAs in TMDs.

These advantages suggest that REAs embedded in 2D materials made of TMDs might be vastly superior to GaAs spin qubits and NV centers in diamond and could pave the way to realizing scalable quantum networks, scalable quantum computing, and ultrasensitive remote quantum sensing. 

%\appendix

\section{References}

%\nocite{*}
\bibliography{bibliography.bib}
\end{document}